\documentstyle[12pt]{article}
\newcommand{\xis}{\raisebox{2.2pt}{$\chi$}}
\newcommand{\negri}{\mbox{\boldmath $\cdot$}}
\newcommand{\Le}{{\cal L}}

\newcommand{\U}{{\cal U}}

\newcommand{\Cael}{{\cal C\ell}}

\newcommand{\om}{\omega}
\newcommand{\infi}{\infty}

\newcommand{\bb}{\begin{equation}}
\newcommand{\ee}{\end{equation}}
\newcommand{\bega}{\begin{eqnarray}}
\newcommand{\ega}{\end{eqnarray}}
\newcommand{\begae}{\begin{eqnarray*}}
\newcommand{\egae}{\end{eqnarray*}}
\newcommand{\nab}{\nabla}
\newcommand{\ga}{\gamma}

\newcommand{\sig}{\sigma}
\newcommand{\sub}{\subset}
\newcommand{\var}{\varphi}
\newcommand{\rig}{\rightarrow}
\newcommand{\longr}{\longrightarrow}

\newcommand{\Longr}{\Longrightarrow}

\newcommand{\al}{\alpha}

\newcommand{\vare}{\varepsilon}
\newcommand{\Ha}{\mbox{\boldmath $H$}}

\newcommand{\La}{\Lambda}
\newcommand{\hb}{\hbar}
\newcommand{\la}{\lambda}

\newcommand{\lan}{\langle}
\newcommand{\de}{\delta}
\newcommand{\ran}{\rangle}

\newcommand{\te}{\theta}

\newcommand{\R}{I\!\!R}

\newcommand{\h}{\hspace*{0.5 cm}}
\newcommand{\dis}{\displaystyle}

\newcommand{\wideF}{\widehat{F}}
\newcommand{\wideg}{\widehat{g}}

\newcommand{\widei}{\widehat{i}}

\newcommand{\tm}{\times}

\newcommand{\sta}{\stackrel}

\newcommand{\ot}{\otimes}
\newcommand{\op}{\oplus}

\newcommand{\cent}{\centerline}
\newcommand{\vs}{\vspace*}
\newcommand{\wede}{{\scriptstyle\wedge}}

\newcommand{\r}{\rho}

\newcommand{\pa}{\partial}
\newcommand{\ma}{\longmapsto}

\newcommand{\E}{\mbox{\boldmath $E$}}
\newcommand{\pp}{\mbox{\boldmath $p$}}
\newcommand{\el}{\mbox{\boldmath $\ell$}}
\newcommand{\J}{\mbox{\boldmath $J$}}
\newcommand{\A}{\mbox{\boldmath $A$}}
\newcommand{\rr}{\mbox{\boldmath $r$}}
\newcommand{\dd}{\mbox{\boldmath $d$}}

\newcommand{\Sa}{\mbox{\boldmath $S$}}
\newcommand{\za}{\mbox{\boldmath $z$}}
\newcommand{\Lu}{\mbox{\boldmath $L$}}

\begin{document}
 
\cent{\Large\bf A Generalization of Dirac Non--Linear}
\cent{\Large\bf Electrodynamics, and Spinning Charged Particles$^{(*)}$}
\footnotetext{$^{(*)}$ Work partially supported by INFN, CNR, MURST and 
by CNPq, CAPES.} 
 
\vs{1.2cm} 

\cent{{Waldyr A. Rodrigues Jr., \ Jayme Vaz Jr.}}

\vs{0.5cm}
 
\h\h\h {\em Departamento de Matem\'{a}tica Aplicada,\\
\h\h\h\h Universidade Estadual de Campinas, 13083--Campinas, S.P.,
Brazil}.\\
 
\

\cent{and}

\

\cent{{Erasmo Recami}}

\vs{0.5 cm}

\h\h\h {\em Dipartimento di Fisica, Universit\`{a} statale di
Bergamo, Bergamo, Italy;\\
\h\h\h\h I.N.F.N., Sezione di Milano, Milan, Italy; and\\
C.C.S., UNICAMP, Campinas, S.P., Brazil.} 

\vs{2cm}

{\bf ABSTRACT:} \ In this note ---dedicated to Prof. Asim O. Barut--- we 
generalize the Dirac {\em non-linear electrodynamics}, by
introducing two potentials (namely, the vector potential $A$ and the 
pseudo-vector potential 
$\gamma^{5} B$ of the electromagnetic theory {\em with charges and 
magnetic monopoles}) \ and 
by imposing the pseudoscalar part of the product 
$\om\om^*$ to be zero, with $\om \equiv A+\ga^5 B$.\\
We show that the field equations of such a theory possess a soliton--like
solution  which can  represent a priori a ``charged particle", since it 
is endowed with a Coulomb field plus the
field of a magnetic {\it dipole}. The rest energy of the
soliton is finite, and the angular momentum stored in its
electromagnetic field can be identified ---for suitable choices of the 
parameters--- with the spin of the charged particle.
Thus this approach  seems to yield
a classical model for the charged (spinning) particle, which does not meet the
problems met by earlier attempts in the same direction. 

\vfill

\newpage 

{\bf 1. INTRODUCTION}\\

\h This note is dedicated to Asim O. Barut, who ---with his inquiring work
about the structure of elementary particles$^{[1]}$--- stimulated so much
researches about the electron structure.
\h As is well-known, classical electrodynamics is based on Maxwell
equations for the electromagnetic field and the Lorentz equations of
motion for the electric charges. This scheme, based on the assumption
that all electromagnetic phenomena be due to existence and motion
of individual electric charges, was sometimes referred to as ``theory of
electrons".$^{[2,3]}$  As pointed out by Whittaker$^{[2]}$, however, that 
name has little  to do with Thomson's discovery
(that for all electrons the ratio $e/m$ is the same); in fact, the 
{\it atomicity} of the
electric charge has little importance in electromagnetism, when
compared with the question of the very {\em structure} of fundamental
charged particles, as the electron.\\
\h This  question was discussed by several authors. The models of
Abraham$^{[4]}$ and Lorentz$^{[5]}$ assumed the electron to be a
spherical, rigid object. Their aim  was to establish a purely
electromagnetic model for the electron. Although much effort has
been made in that direction, those models met difficulties and problems, 
which are
well discussed in the literature.$^{[2,3,6]}$ \ In 1938 Dirac$^{[7]}$
abandoned the idea of looking for the electron structure,
and tried to develop on the contrary a classical theory for a 
{\it point--like} electron. The point--like
electron,  however,  exhibited  an infinite self-energy: as
expected. \ Therefore, no conclusive and satisfactory solution was given
---nor seems to have been given--- to the problem of the structure of charged 
particles (unless one abandons differential equations for finite--difference
equations, following Caldirola$^{[8]}$).\\
\h In order to overcome the difficulties associated with point charges, 
Dirac$^{[9]}$ suggested in 1951 a trick, which allows describing 
electric currents when starting from a theory ``without charges". Namely, Dirac 
{\em exploited} the presence in electromagnetism of a vector
potential $A$ ({\em i.e.}, of those extra variables entering the 
electromagnetic theory because of its gauge invariance); imposed
a non-linear gauge: i.e., the condition that the potential 1-form $A$ 
were  timelike [$A^2$ = $k^{2}$ = positive constant]; so that $A$ could be 
regarded as proportional to a velocity field$^{[10]}$; and finally identified 
it with a current,$^{[11]}$ 
getting the motion of a continuous stream of electricity (rather than
the motion of point charges) . This theory is known as Dirac 
non-linear electrodynamics.$^{\#1}$\\
\footnotetext{$^{\#1}$ It is noteworthy that already Schroedinger$^{[11]}$  
mentioned two cases in which the potential did actually acquire the role
of source of the very field that was derived from it: (i) the case of 
the scalar potential in Debye's theory of electrolytes; and  (ii) of  
the vector potential in London's theory of superconductivity.}
\h Recently, Righi and Venturi have shown in an interesting
paper$^{[12]}$ that the field equations of Dirac's new theory admits an 
extended--type,
spherically symmetric, static solution which can be considered as a
charged particle. In their approach, however, a magnetic dipole moment 
(and a spin) can arise only as a first-order {\em quantum} effect; but
this would imply the electromagnetic potential to be no longer 
(at the second order) a 1-form with constant square: contrarily to the 
initial assumption. Moreover, the classical radius $r_{0}$ is given
in ref.$^{[12]}$ by an equation of the form $f(r_0)=0$, with the condition
 \ $\cosh f(r)\leq 1$: \ which means that $A$ has to assume imaginary
values.\\ 
\h In this paper we shall present an alternative approach to the
question of the structure of charged particles that, although similar to
Dirac's and Righi--Venturi's, does not meet the kind
of problems discussed above. In fact, instead of the usual Maxwell equations,
let us introduce the {\em generalized} Maxwell equations, including magnetic
monopoles.  We shall be able, then, to produce a new formulation of
Dirac's  non-linear electrodynamics, in which 
the first ({\em vector}) potential $A$ is orthogonal to the
second ({\em pseudo-vector}) potential $\star B \equiv -\gamma^{5} B$, 
symbol $\star$ representing ---in the language of differential forms--- 
the Hodge dual. Thus, we shall obtain 
the Righi--Venturi results in a purely classical way (i.e., without 
any need of introducing spin as a quantum effect).\\ 
\h To introduce our own approach, in Sect. {\bf 2} we reformulate
Maxwell equations,  and their generalized version, in terms of the Clifford
bundle formalism, which provides us with a natural and concise formulation 
of them.  \ In Sect. {\bf 3} we present our approach,  and in Sect. {\bf 4} 
we look for solutions of our field equations. A relevant solution
is constituted by the electric
field of a conducting sphere and by the magnetic field 
of a magnetized sphere; while the angular momentum stored in
the total electromagnetic field is equal to $\hb/2$. \ In Sect. {\bf 5} 
we discuss this solution.\\
 
\vs{5mm}
 
{\bf 2. THE CLIFFORD--BUNDLE APPROACH TO
ELEC\-TRO\-MAG\-NE\-TISM$^{[13,14,15]}$}\\
 
\h Let $M$ denote the Minkowski spacetime  and $T^*M \; [TM]$ the
cotangent [tangent] bundle of $M$. Cross-sections $e \in \sec TM \; [\ga
\in \sec T^*M]$ will be called 1-form [1-vector] fields. Let $g \in
\sec (T^*M \tm T^*M) \; [\wideg \in sec (TM \tm TM)]$ such that in each
fiber $\pi^{-1}(x), \; \; x \in M , \;$ quantity $g_x \; [\wideg_x]$ is a 
symmetric bilinear form over
$T_xM \; [T^*_xM]$. Let $\{e_\mu\} \in \sec TM$ be a basis of $TM$ and
$\{\ga^\mu\} \in \sec T^*M$ be the dual basis. Then $g =
\eta_{\nu\mu}\ga^\mu \ot \ga^\nu \; \: [\wideg = \eta^{\mu\nu} e_\mu \ot
e_\nu]$, where $(\eta^{\mu\nu}) =$ diag($1,-1,-1,-1$).\\
\h Let $T(V) = (\op^\infi_{p=0} \ot^p V, \op)$
be the tensor algebra over the real field, where $\ot$ is the usual tensor
product, $\op$ denotes the weak direct sum and $\ot^pV$ the cartesian
product of $p$ copies of $V$. The Clifford algebra is the quotient
algebra $T(V)/J$, where $J$ is the two-sided ideal generated by
elements of the form $a \ot a - Q(a)$ with $a \in V$ and 
$Q(a) = g(a,a)$. A vector bundle is called a Clifford bundle if
each fiber is a Clifford algebra. For more details see$^{[13-15]}$.\\
\h Let $\Cael(M, \wideg)$ be now the Clifford bundle {\em of differential 
forms} over
Minkowski spacetime. The spacetime algebra $\R_{1,3}$ is the  typical
fiber of the Clifford bundles over Lorentzian space-times. The Dirac 
operator $\pa$ acting on
sections of $\Cael(M,\wideg)$ is $\pa = \ga^\mu\nab_\mu$, where
$\nab$ is the Levi-Civita connection of $g$. We can choose for
simplicity $\{\ga^\mu\}$ such that $\nab_\mu = \pa_\mu$; thus $\pa =
\ga^\mu\pa_\mu$. We also have $\pa=d-\de$, where $d$ is the
differential and $\de$ the Hodge codifferential operator.\\
\h The Maxwell equations are
\bb
dF = 0 \;\;;\;\; \de F = - \frac{4\pi}{c}\,J \,,
\ee
where the electromagnetic field $F \in \sec \La^2(M) \sub \sec
\Cael(M, \wideg)$ is a 2-form field \linebreak $(F \equiv \frac{1}{2}
F_{\mu\nu}\ga^\mu \wede \ga^\nu)$ and the electric current
$J \in \sec \La^1(M) \sub \sec \Cael(M,\wideg)$ is a 1-form field
$(J=J_\mu\ga^\mu)$. In the Clifford bundle the Maxwell equations (1)
can be written, by using the Dirac operator $\pa = d-\de$, as
\bb
\pa F = \frac{4\pi}{c}\,J \,,
\ee
which was originally due to Riesz.$^{[16,17]}$ \ If we denote by
$\negri$ the internal product and by $\wede$ the external product,
we have $\pa F = \pa\negri F + \pa \wede F$, so that $\pa\wede=d$ and 
$\pa\negri=-\de$. In terms of the electromagnetic potential
$A \in \sec \La^1(M) \sub \sec \Cael(M,\wideg)$  we have
\bb
F = \pa \wede A \; 
\ee
or, in the Lorenz$^{(*)}$
\footnotetext{$^{(*)}$ It is perhaps time to recognize that the gauge
condition  $\pa\negri A=0$ is due to Lorenz and not to
Lorentz.} gauge $\pa\negri A=0$,
\bb
F=\pa A \,,
\ee
so that $\pa^2 A = \Box A = \frac{4\pi}{c}\,J$ is the wave equation,
where $\Box = \pa^2 = (d-\de)^2 = -(d\de+\de d)$ is the
Laplace-Beltrami operator.\\
\h Let the main anti-automorphism in $\R_{1,3}$ (called reversion) be
denoted by $*$, i.e.,\linebreak  
$(AB)^* = B^* A^*$ and  $A^*=A$ if $A$ is a
scalar or a 1-form. Since $F$ is a 2-form we have $F^*=-F$. Now it is
easy to show that
\bb
\pa_\mu S^\mu  = \pa_\mu(-\frac{c}{8\pi}\,F\ga^\mu F)=J\negri F\,,
\ee
where $J\negri F = \frac{1}{2}(JF-FJ)$. We call the quantities
$S^\mu$ the energy momentum 1-forms; then
$E^{\mu\nu}=S^\mu\negri\ga^\nu$ are the components of the
energy-momentum tensor. In fact, if we project the elements of
$\R_{1,3}$ into the Pauli algebra $\R_{3,0}$ with
$\sig^i=\ga^i\ga^0\;(i=1,2,3)$, we have that $F=\E+\widei\Ha$, where
$\E=E_i\sig^i\;,\; \Ha=H_i\sig^i\;,\; \widei=\sig^1\sig^2\sig^3$ and
$E_i=F_{0i}\;,\; H_1=F_{23}\;,\;$\linebreak
$H_2=F_{31}\;,\;H_3=F_{12}\;$, and that
$J\ga^0=c\r+\J\;,\; S^0\ga^0=c \: \U+\Sa^0$; \  and we obtain
\bb
S^0\ga^0=c \: \U+\Sa^0=\frac{c}{8\pi}(\E^2+\Ha^2)+\frac{c}{4\pi}\E\tm \Ha
\ee
where we recognized
\bb
\U \equiv \frac{1}{8\pi}(\E^2+\Ha^2)\;\;\;,\;\;\; \Sa^0 \equiv 
\frac{c}{4\pi }\E\tm \Ha
\ee
as the energy density and the Poynting vector of the electromagnetic
field, respectively (in expression (6) we also used $\E\tm\Ha = -\widei\E 
\wede \Ha$). As well-known
\bb
\pp = \frac{\Sa^0}{c^2} = \frac{1}{4\pi c} \E \tm \Ha\;\;;\;\; \el = \rr 
\tm \pp
\ee
are the electromagnetic momentum density and the angular
momentum density, respectively.\\
\h It is important to notice that with the Clifford bundle approach the
Lorentz force equation of motion need  not to be postulate. In fact, let us
write $-K =J\negri F$ and project it onto the Pauli algebra, so
that
\bb
K\ga^0 = \J\negri\E + c(\r\E+\J\tm\Ha)\,.
\ee
If we define $\pa_\mu M^{\mu\nu} = K\negri\ga^\nu$ we have from eq.(5)
that
\bb
\pa_\mu(E^{\mu\nu} + M^{\mu\nu})=0\,,
\ee
where $M^{\mu\nu}$ plays the role of the symmetric energy-momentum
tensor of matter (i.e., of the currents) in the above global
conservation equation. After this crucial identification, and following 
Barut,$^{[19]}$ we may write (where $m$ appears for dimensional reasons):
\bb
\begin{array}{l}
M^{\mu\nu} = -m \int ds\,\de(x^\al-z^\al){\dis\frac{dz^\mu}{ds}}
\,{\dis\frac{dz^\nu}{ds}} \; ,\\
\\
J^\mu = e \int ds\,\de(x^\al-z^\al){\dis\frac{dz^\mu}{ds}}\, ,
\end{array}
\ee
which give the correct equations of motion, i.e.:
\bb
m\ddot{\za}_i = e [\E_i + \frac{1}{c}(\dot{\za}\tm\Ha)]_i\,.
\ee
\h Now, the Clifford bundle formalism can be used in order to
formulate the Maxwell equations with magnetic
monopoles.$^{[14,15]}$ Namely, Maxwell equations can be generalized,
in order  to include monopoles (without string!),$^{[15]}$ in the form:
\bb
dF=-\frac{4\pi}{c} \star J_m \;\;\;;\;\;\; \de F = -\frac{4\pi}{c}J_e
\ee
where $J_e$ and $J_m$ are the electric and magnetic
currents (with $J_e \neq k J_m$), respectively, and $\star$ denotes the  Hodge
star operator. Notice that in the present way {\em electric} and {\em magnetic}
 charges  have been introduced as objects of similar nature;$^{[14]}$  in 
particular, monopoles are {\em not} associated to strings, and the topology of
Minkowski space-time has not been modified. \ Then, by
using the Dirac operator $\pa=d-\de$ and the fact that 
$\star J =  J^*\ga^5$, in the Clifford bundle we have
\bb
\pa F=\frac{4\pi}{c} \; J
\ee
where
\bb
J=J_e + \ga^5 J_m \,.
\ee
By introducing the electromagnetic pseudo-potential $\ga^5 B$ of
Cabibbo and Ferrari,$^{[20]}$ where $B \in \sec \La^1(M) \sub \sec
\Cael(M,\wideg)$, we can write
\bb
F=F_e+\ga^5 F_m
\ee
with
\bb
F_e \equiv \pa\wede A\;\;\;,\;\;\; F_m \equiv -\pa\wede B\,.
\ee
And, in the Lorenz gauge, $\pa\negri A=0,\; \pa\negri B=0$, we have:
\bb
F=\pa A - \ga^5\pa B = \pa(A+\ga^5 B) = \pa\om
\ee
where $\om \equiv A+\ga^5 B$ is the generalized electromagnetic
potential.$^{[14]}$\\
\h From the generalized Maxwell equations it follows that 
\bb
\pa_\mu S^\mu=\pa_\mu\biggl(-\frac{c}{8\pi} F\ga^\mu F\biggr)
=J_e\negri F + J_m \negri(\ga^5 F)\,,
\ee
where $S^\mu$ are the energy-momentum 1-forms, such that
$S^0\ga^0=c \: \U+\Sa^0$ with $\U$ and $\Sa^0$ given by eq.(7).
In$^{[14,15]}$ we showed  that $-K_e = J_e\negri F$ and
$-K_m=J_m\negri(\ga^5 F)$ correspond to the expressions for the
Lorentz (electric and magnetic) forces, and by writing\linebreak 
$\pa_\mu M^{\mu\nu}=K_e\negri\ga^\nu+K_m\negri\ga^\nu$ we succeeded in
obtaining the correct equations of motion.\\
\h Let us stress that the generalized Maxwell equations can be
obtained either from the ``non canonical" lagrangian adopted by us in 
refs.$^{[14]}$ (which contained both a scalar and a pseudoscalar part), or from 
the following lagrangian$^{[15]}$
\bb
\Le=\frac{1}{8\pi} \lan F \wideF\ran_0 - \frac{1}{c} \lan J\om
\ran_0
\ee
where the brackets indicate the scalar part, and
\bb
\wideF \equiv  F_e -\ga^5 F_m \, \; .
\ee
 
\vs{5mm}
 
{\bf 3. AN ALTERNATIVE APPROACH TO FREE ELECTROMAGNETISM}\\
\h We have recalled in Sect. {\bf 1} that Dirac's 1951 theory$^{[10]}$ does not
seem to be able to describe a (spin ${1 \over 2}$) electric particle endowed
with a magnetic dipole field ---besides the electrostatic one---; unless
we start, as we are going to see, from Maxwell equations with monopoles.\\   
\h Let us consider, from now on, the {\em generalized} electromagnetic field 
in {\em free space}, so that $J=0$, i.e., $J_e=J_m=0$. In this case, the free
generalized Maxwell equations
\bb
\pa F = 0
\ee
can be obtained from the lagrangian 
\bb
\Le=\frac{1}{8\pi} \lan F \wideF \ran_0 \,.
\ee
We get the {\em usual} free Maxwell equations by taking the second potential 
equal to zero, $B=0$,  so that
$F_m=-\pa\wede B=0$ and
\bb
\pa F_e \equiv 0 \,.
\ee
The following scheme summarizes this approach:
\bb
\Le=\frac{1}{8\pi}\lan F \wideF \ran_0 \;\;\;
\sta{{\rm
Euler-Lagrange}}{_{\sta{--------------\longrightarrow}{{\rm equation}}}}
\;\;\; \pa F = 0 \,
\sta{B=0}{\longr}\; \pa F_e = 0\;.
\ee
\h Clearly the condition $B=0$ can be imposed from the beginning, but
in this case the lagrangian has to be modified. Let us introduce a
Lagrange multiplier $\la=\la(x)$ so that our lagrangian turns into 
\bb
\Le=\frac{1}{8\pi}\lan F \wideF \ran_0 +\frac{1}{2c} \la(x) B^2\,.
\ee
Now $\la(x)$ must be considered as a field quantity, so that the
Euler-Lagrange equations for $A(x), \, B(x)$ and $\la(x)$ are
\bb
\pa F_e = 0
\ee
\bb
\pa F_m=-\frac{4\pi}{c}\,\la B
\ee
\bb
B^2  = 0 \;.
\ee
The new equation (28) implies that $\pa\wede F_m=0$ and $\pa\negri F_m =
-\frac{4\pi}{c}\,\la B$, so that from the latter we have $\pa\negri
(\la B)=0$, which is just the continuity equation $\pa\negri J=0$ with 
$J=-\la B$. But eq.(28) tells us that
$\pa\negri (\la B) = (B\negri\pa)\la + \la(\pa\negri B) = 0$ and,
since in general $\pa_\mu\la\neq 0$, if from eq.(29) we have as
solution $B\neq 0$ we must have $\pa\negri B\neq 0$. However, it is always
possible to find a gauge transformation such that $\pa\negri B=0$
(Lorenz gauge), and then we must have $B=0$ (provided that in general $\pa_\mu
\la\neq 0$) and $F_m=-\pa\wede B=0$. \ In other words, it seems a priori that
in eq.(29) one may have $B \neq 0$; but eq.(28) implies that $B = 0$. \ 
Thus the usual free Maxwell
equations  are recovered according to the following scheme:
\bb
\Le=\frac{1}{8\pi}\lan F \wideF \ran_0
\sta{B=0}{\longr}\;
\Le=\frac{1}{8\pi}\lan F \wideF \ran_0  + \frac{1}{2c}\,\la B^2
\;\;\sta{{\rm
Euler-Lagrange}}{_{\sta{--------------\longrightarrow}{{\rm equation}}}}
\;\;\; \pa F_e = 0 \,.
\ee
\h Now let us consider the generalized potential $\om$ given by
eq.(18), \ $\om \equiv A+\ga^5 B$. Let us calculate $\om\om^*$:
\bb
\om\om^*=(A+\ga^5 B)(A-\ga^5 B) = (A^2-B^2)+2\ga^5(A\negri B)\,,
\ee
so that its scalar and pseudo-scalar parts are
\bega
&& \lan \om\om^* \ran_0 = A^2 - B^2 \,,\\
&& \nonumber\\
&& \lan \om\om^* \ran_4 = 2\ga^5(A\negri B)\,,
\ega
respectively. If $B=0$, we have
$\lan \om\om^* \ran_0 = A^2$ and $\lan \om\om^* \ran_4 = 0$, i.e. \ 
the pseudo-scalar part of $\om\om^*$ vanishes. Thus
\bb
B=0 \Longr \lan \om\om^* \ran_4 = 0
\ee
and
\bb
B=0 \Longr \om\om^* = s
\ee
where the quantity $s$ is a {\em scalar}, not necessarily
a constant; i.e.:
\bb
s=s(x) \in \sec \La^0(M) \sub \sec \Cael(M,\wideg)\,.
\ee
Clearly the condition $\lan \om\om^* \ran_4 = 0$ is {\em more
general} than $B=0$, so that $B=0$ follows as a particular
case of $\lan \om\om^* \ran_4 = 0$ or $\om\om^*=s$.
What we shall propose here is to {\em replace the condition $B=0$ by
the more general condition} $\om\om^*=s$, thus extending the Dirac original,
non-linear gauge condition $A^{2} = k^{2}$. \  By eq.(31) the condition
$\om\om^*=s$ implies the two following constraints:
\bb
A^2 - B^2 = s
\ee
\bb
A\negri B = 0 \; .
\ee
Notice that we do not want {\em here} to modify the lagrangian 
$\Le = \frac{1}{8\pi}\lan
F \wideF \ran_0$ by adding a term of the type $\la(\om\om^* - s)$ because it
contains both a scalar and a pseudo-scalar, and this may be regarded as 
unconventional. A possible approach consists in using two
Lagrange multipliers, \ $\lambda_{1}(x)$ and $\lambda_{2}(x)$, \  one for 
each condition (37) and (38).
Thus our lagrangian will be
\bb
\Le = \frac{1}{8\pi}\lan F \wideF \ran_0 + \frac{\la_1}{2c}
(A^2-B^2-s) + \frac{\la_2}{c}(A\negri B)
\ee
or, since $\lan F \wideF \ran_0  = \lan F^2_e + F^2_m \ran_0$, 
we shall have, in terms of components:
\bega
\Le&=&-\frac{1}{16\pi}[(F_e)_{\mu\nu}(F_e)^{\mu\nu}
+(F_m)_{\mu\nu}(F_m)^{\mu\nu}]+ \nonumber\\
&&\nonumber\\
&+&\frac{\la_1}{2c}[A_\mu A^\mu - B_\mu B^\mu -s] +
\frac{\la_2}{c}(A_\mu B^\mu)\,.
\ega
Our new field equations (which generalize the Dirac non-linear theory of 
electromagnetism) are, in components, the following:
\bega
&&\pa_\mu(F_e)^{\mu\nu} = - \frac{4\pi}{c} \la_1 A^\mu - \frac{4\pi}{c}
\la_2 B^\mu\\
&& \nonumber\\
&&\pa_\mu(F_m)^{\mu\nu} = \frac{4\pi}{c} \la_1 B^\mu - \frac{4\pi}{c}
\la_2 A^\mu
\ega
\bb
A_\mu A^\mu - B_\mu B^\mu = s
\ee
\bb
A_\mu B^\mu = 0
\ee
or, in intrinsic form:
\bega
&&\pa F_e = \frac{4\pi}{c} \la_1 A + \frac{4\pi}{c}\la_2 B\\
&& \nonumber\\
&&\pa F_m = -\frac{4\pi}{c} \la_1 B + \frac{4\pi}{c}\la_2 A
\ega
\bb
A^2 - B^2 = s
\ee
\bb
A \negri B = 0 \,.
\ee
From eq.(45) and eq.(46) it follows that our new field equations can be simply
written down as 
\[
\pa F = \frac{4\pi}{c} \la_1 \om - \frac{4\pi}{c}\la_2\ga^5\om =
\frac{4\pi}{c}(\la_1-\la_2\ga^5)\om \; ,
\]

and therefore:
\bb
\fbox{$\pa F = {\dis\frac{4\pi}{c}} \; \la\om\,\,$} \; ,
\ee
with  $\la = \la_1 - \la_2 \ga^5\,.$\\
\h Furthermore, let us impose the gauge invariance of our theory. For
$A \ma A' = A + \pa \xis_1$ and $B \ma B' = B = + \pa \xis_2$,  and from 
eqs.(45) and (46), in order for equations (49) to be gauge invariant
we must have the following constraints on the Lagrange multipliers: 
\bega
&& \la_1\pa\xis_1 + \la_2\pa\xis_2 = 0\\
&&\nonumber\\
&& \la_2\pa\xis_1 + \la_1\pa\xis_2 = 0 \; .
\ega
 
\vs{5mm}
 
{\bf 4. A SOLUTION OF THE NEW FIELD EQUATIONS}\\
 
\h Now let us look for solutions of our new field equations, 
corresponding to a charged particle at rest.  In other words, 
let us here consider only {\em stationary} solutions.
First we note that the condition $A\negri B=0$ is trivially satisfied
by $B=0$. Since we want this condition to be gauge invariant, i.e.
$\lan \om'\om'^* \ran_4 = \lan \om\om^* \ran_4 = 0$, we must have
\bb
\lan \om'\om'^* \ran_4 = 2\ga^5(A' \negri B') = 0
\ee
and, since $B' = \pa\xis$ (because $B=0$),
\bb
A' \negri \pa\xis_2 = 0 \,.
\ee
But $B=0$ implies, in eq.(46), that
\bb
\la_2 A = 0
\ee
and (since we supposed $A\neq 0$) then
\bb
\la_2 = 0 \,.
\ee
The gauge invariance of our theory is guaranteed, from eq.(50) and
eq.(51), once
\bega
&& \la_1\pa\xis_1 = 0\\
&&\nonumber\\
&&\la_1\pa\xis_2 = 0 \,.
\ega
We now look for a $\xis_2 = \xis_{2}(x)$ that does not depend on $x^0$ = ct
and is spherically symmetric, i.e.:
\bb
\xis_2 = \xis_{2}(r)
\ee
where $r=|\rr|$. Thus we have
\bb
\pa_0\xis_2 = 0 \;,\; \pa_i\xis_2 = \frac{d\xis_{2}(r)}{dr}\frac{x_i}{r}
\equiv \var(r)\frac{x_i}{r}
\ee
and, for $A'=A_0\ga^0+A_i\ga^i$, the condition (53) gives:
\bb
A'\negri\pa\xis_{2} = -A_i\var(r)\frac{x_i}{r} =
-\A\negri\big[\var(r)\frac{\rr}{r}\big] = 0 \; ,
\ee
where $\A=A_i\sig^i\;,\; \rr = x_i\sig^i$ and $\sig^i=\ga^i\ga^0, \;
i=1,2,3$. Note that, since
$\pa_0\xis_{2} = 0$, the choice of $A_0$ is arbitrary; we shall suppose that
$A_0(x)$ does not depend on $x^0$ and is spherically symmetric,
i.e.:
\bb
A_0(x) = A_0(r)\,.
\ee
From eq.(60), $\A$ must be orthogonal to $\rr$. Since
$(\dd\tm\rr)\negri\rr=0$ for any vector $\dd$, we take
\bb
\A \equiv \phi(r)(\dd\tm\rr) \; ,
\ee
where $\dd$ is a vector to be defined, and $\phi(r)$ a function to be
calculated (just defined via eq.(62)).  In summary, our potentials $A$ and $B$
are$^{(*)}$
\footnotetext{$^{(*)}$ From now onward, we shall omit the prime
and write $A$ and $B$ (since there is no danger of confusion).}
 
\vs{0.5cm}
 
$$
A=A_0\ga^0+A_i\ga^i \qquad {\rm with} \qquad \left\{\begin{array}{l}
A_0=A_0(r)   \\  A_i=\phi(r)(\dd\tm\rr)_i \end{array}  \right.
\eqno{\begin{array}{c} {\rm (64)} \\ {\rm (65)} \end{array}}$$
 
and
 
$$B=B_0\ga^0+B_i\ga^i \qquad {\rm with} \qquad \left\{\begin{array}{l}
B_0=0  \\  B_i=\var(r){\dis\frac{x_i}{r}} \; . \end{array} \right.
\eqno{\begin{array}{c} {\rm (66)} \\ {\rm (67)}\end{array}} $$
 
\vs{0.5cm}
 
\h Finally we notice that condition (57) for gauge invariance yields
$\la_1\pa_\mu\xis=0$, and from eq.(59) we must have $(r\neq 0)$:
\setcounter{equation}{67}
\bb
\la_1\var(r)=0\,.
\ee
For $r$ such that $\var(r)\neq 0$ we must then have $\la_1=0$; if we
suppose that there exists a value of $r$ (namely $r_0$) such that
\bb
\var(r_0)=0
\ee
we get that $\la_1$ must be of the form
\bb
\la_1=-\frac{\de(r-r_0)}{r_0}\; \frac{c}{4\pi} \; ,
\ee
where $r_0$ in the denominator has been introduced for dimensional reasons,
and the other constants (and the minus sign) for convenience. The other
condition (56) implies that $\xis_{1} = \xis_{1}(x)$ is an arbitrary function,
except at $r=r_0$ where $\pa{\xis_{1}}|_{r_0}=0$. Thus our theory is gauge
invariant except at $r=r_0$.$\: ^{(*)}$\\ \footnotetext{$^{(*)}$ Notice
also that $\var(r_0)=0$ is an extremum condition since from eq.(60)
we have $\var(r_0)=\frac{{\rm d}\xis_{2}(r)}{{\rm d}r}|_{r_0} = 0$.}
\h At this point, we can claim that eqs.(46)-(48) are satisfied by the above
choices of the potentials and Lagrange multipliers. The remaining
eq.(45) will specify $A_0$ and $A_i$. Thus we have
\bb
\pa_\mu F_e^{\mu\nu} = -\frac{\de(r-r_0)}{r_0}\,A^\nu \; .
\ee
Remembering that $F^{i0} = -F_{i0}\,,\, F^{ij}=F_{ji}\;\;
(i,j=1,2,3\,,\,i\neq j)$ and using eq.(63) and eq.(64), we obtain for
$F_e^{\mu\nu} = \pa^\mu A^\nu - \pa^\nu A^\mu\;(i\neq j)$ that:
\bega
&& F^{i0}_e = -\frac{d A_0(r)}{dr} \;\frac{x_i}{r}\\
&&\nonumber\\
&& F^{ij}_e=\frac{d \phi(r)}{dr} \;\frac{x_i}{r}(\dd\tm\rr)_j
-\frac{d \phi(r)}{dr} \;\frac{x_i}{r}(\dd\tm\rr)_i +
\vare_{ijk}2\phi(r)d_k\,.
\ega
Then, for $\nu=0$, eq.(71) gives:
\bb
\frac{d^2A_0(r)}{dr^2}+\frac{2}{r}\,\frac{dA_0(r)}{dr} =
\frac{\de(r-r_0)}{r_0}\,A _0(r)
\ee
and, for $\nu=1,2,3$, the same eq.(71) gives the equation 
\bb
\frac{d^2\phi(r)}{dr^2}+\frac{4}{r}\,\frac{d\phi(r)}{dr} = -
\frac{\de(r-r_0)}{r_0}\,\phi(r)\,.
\ee
A solution of eq.(74) is of the form
\bb
A_0(r) = \frac{c_1}{r} \Theta(r-r_0)
\ee
where $c_1$ is a constant and $\Theta$ is the step function; that is:
\bb
A_0(r)=\left\{\begin{array}{cl}
{\dis\frac{c_1}{r}} \;\; , \qquad & r \geq r_0\\
&\\
0 \;\; , \qquad & r < r_0 \; ; \end{array}\right.
\ee
while a solution of eq.(75) is of the form
\bb
\phi(r) = \left\{\begin{array}{cl}
{\dis\frac{c_2}{r^3}} \;\; , \qquad & r > r_0\\
&\\
0 \;\; , \qquad & r \leq r_0 \;, \end{array}\right.
\ee
which we can simply write (by using the step--function $\Theta$):
\bb
\phi(r) = \frac{c_2}{r^3} \Theta (r-r_0) \; ,
\ee
with, moreover, $\phi(r_0)=0$ for $r=r_0$. \ From eq.(65) we get therefore:
\bb
A_i = \frac{c_2}{r^3} (\dd\tm\rr)_i \Theta (r-r_0)\,.
\ee
\h Now, using the last results in eq.(72) and eq.(73), we obtain
\bb
F^{i0}_e = \left\{\begin{array}{cl}
0 \;\; , \qquad & r < r_0 \\
&\\
c_1{\dis\frac{x_i}{r^3}} \;\; , \qquad & r \geq r_0
\; \end{array}\right.
\ee
 
\vs{2mm}
and
\vs{2mm}

\bb
F^{ij}_e = \left\{\begin{array}{cl}
0 \;\; , \qquad & r \leq r_0 \\
&\\
{\dis\frac{3c_2}{r^5}}
[x_j(\dd\tm\rr)_i-x_i(\dd\tm\rr)_j]+2\vare_{ijk}{\dis\frac{c_2}{r^3}}
d_k \;\; , \qquad & r > r_0
\; . \end{array}\right.
\ee
Remembering that $F^{i0}_e = E^i$ and $F^{ij}_e = H^k \; (i,j,k$
cyclic); choosing $\dd$ parallel to $\sig^3$; and using spherical 
coordinates, we eventually get the soliton--like solution
\bb
\E = \left\{\begin{array}{cl}
0 \;\; , \qquad & r < r_0 \\
&\\
c_1{\dis\frac{\widehat{\rr}}{r^3}} \;\; , \qquad & r \geq r_0
\;; \end{array}\right.
\ee
 
\vs{5mm}
 
\bb
\Ha = \left\{\begin{array}{cl}
0 \;\; , \qquad & r \leq r_0 \\
&\\
c_2{\dis\frac{1}{r^3}}(2\cos\te\widehat{\rr}+\sin\te \widehat{\te})
\;\; , \qquad & r > r_0
\;. \end{array}\right.
\ee
 
where $\te$ is the angle between $\widehat{\rr}$ and $\sig^3$. The
electric field $\E$ is like the one outside  a conducting, charged  sphere; 
and the
magnetic field $\Ha$ looks like the one outside a magnetized sphere (except
that inside the sphere we have {\em no} magnetic field). Notice that in the
interior of the sphere both $\E$ and $\Ha$ are zero, according to this 
solution.\\
\h The constants $c_1$ and $c_2$ are
proportional to the electric charge and to magnetic moment, respectively. \ 
Indeed from eq.(71), if we identify $\frac{4\pi}{c}\,J^0 =
-\frac{\de(r-r_0)}{r_0}\;A^0$, where $J^0 = c\r$, we have
\bega
\int d^3 x J_0&=&c\,e = -\frac{c}{4\pi}\int d^3 x
\frac{\de(r-r_0)}{r_0}\;\frac{c_1}{r}= \nonumber\\
&&\nonumber\\
&=&-\frac{c}{4\pi} \int dr \, r\de(r-r_0)=-cc_1 \Longr c_1 = -e
\ega
Now, if we choose
\bega
&& r_0 \equiv \frac{e^2}{2mc^2}\biggl[1+\frac{3}{4}
\biggl(\frac{\hb c}{e^2}\biggr)^2\biggr]\\
&& \nonumber\\
&& c_2 \equiv \frac{3}{8}\; \frac{e\hb}{mc}\biggl[1+\frac{3}{4}
\biggl(\frac{\hb c}{e^2}\biggr)^2\biggr]
\ega
we obtain the whole mass $m$ to be electromagnetic in origin:
\bb
\U = \frac{1}{8\pi} \int\;(\E^2 + \Ha^2)dv = mc^2 
\ee
and the angular momentum of the whole Poynting vector field to get the value
\bb
\Lu = \frac{1}{c^2} \int (\rr \tm \Sa^0)dv = \frac{\hb}{2} \sig^3\;,
\ee
so that $|\Lu| = L_z = \hb/2$. \ In connection with eqs.(86), (87), one can 
immediately observe that the magnetic moment $c_2$ vanishes only if 
$\hbar \rig 0$, in which case $r_0$ is just equal to  the so--called 
classical radius of the electron (without spin).  On the contrary,
if our particle {\em has} spin ($\hbar \neq 0$), then $r_0$ jumps$^{(*)}$ 
\footnotetext{$^{(*)}$ {\bf Unless one attributes a large enough value to} 
$m$,
{\bf or replaces in eq.(86) the coupling--constant} $\alpha \equiv {e^{2} /
\hbar c}$ {\bf by the constant} ${g^{2} / \hbar c}$ {\bf (quantity} $g$ 
{\bf being the elementary magnetic--pole charge).}} 
to the value
of the ``electromagnetic world"$^{[21]}$  radius: {\em i.e.}, to a value 
of  the order
of the hydrogen atom size ($1$ \AA). It should be stressed, however, that 
the interior of such a sphere is completely accessible to any other
particle (apart from possible electromagnetic repulsions).

\newpage

{\bf 5. CONCLUSIONS}\\

\h Let us comment again  about the solution we just found.
It yields the model for a charged particle as a ``sphere" with  radius
$r_0= \frac{e^2}{2mc^2}[1+\frac{3}{4}(\frac{\hb c}{e^2})^2]$, which
for $\hb \rig 0$ is just the electron classical radius
$r_{\rm cl}=\frac{e^2}{2mc^2}$. 
The electric field looks just like the one of a
conducting sphere, which is a most natural case if the
electron is to be modelled as a sphere. On the other hand, although
the magnetic field looks like the one of a magnetized sphere, it is
indeed different. For a magnetized sphere the magnetic field is
non-vanishing also inside the sphere
and at its surface, while for our solution it vanishes both inside
the sphere and at its surface. [Actually, even if the magnetic field 
happened to be different
from zero inside the sphere and at its surface, the angular momentum
stored in the field would turn out to be the same, i.e., $L_z
=\hb/2$, since the electric
field does vanish inside the sphere.] \ Therefore, the present solution
suggests for a charged particle the picture of 
a hole in an ``electromagnetic fluid".\\
\h Finally, let us make some comments about our approach. Clearly it
does not meet the same kind of problems discussed in the introduction,
concerning the approach of Righi and Venturi. Moreover, it is
``classical" in two senses: first, that it is ``non-quantistic"; second,
that ---although we started from the generalized Maxwell equations--- we
did assume afterwards the pseudo-potential to be  zero, i.e.
$\ga^5B=0$: which is the case for ordinary  electromagnetism. Thus,
our solution seems to picture a charged particle as a sphere with nothing 
inside it, and such that an angular momentum $\hb/2$ is stored in its field
(even if it has nothing to do with any internal degrees of freedom like spin).
 It may be  interesting to compare this result with the  discussion in
Hestenes$^{[22]}$ concerning the possible meaning of spin and the
``zero-point" angular momentum in quantum mechanics.\\
\h The possible meaning of our solution, and its possible role in the
physical situations in which Dirac's non-linear electrodynamics is an
acceptable theory, may need more discussion.\\ 

\vs{5mm}

{\bf 6. Acknowledgments}\\

Thanks are due, for stimulating discussions, to G. Barbiellini, B. Bertotti
(also for some historical refs.), G. Bonera, A. Insolia, A. Panatta, 
A. Piazzoli,   
G.D. Maccarrone, E. Majorana jr., R. Potenza, F. Raciti, G. Salesi, 
M. Sambataro, S. Sambataro, J.B. Teixeira, E. Tonti, M.T. Vasconselos
and J.R.R. Zeni. \ One of the authors [ER] is moreover grateful to
A. Inomata, R. Wilson and particularly A. van der Merwe for their kind
invitation to contribute to this issue in honor of Professor Asim O. Barut.

\newpage

{\bf 7. REFERENCES}

\begin{list}{ }{\leftmargin 12 mm \labelwidth  7mm \labelsep 1mm}

\item [{[1]}\hfill] See, e.g., A.O. Barut: {\em Lett. Math. Phys.}
{\bf 10} (1985) 195; A.O. Barut and R. R\c{a}czka: {\em Lett. Math. Phys.}
{\bf 1} (1976) 315; A.O. Barut and A. Zanghi: {\em Phys. Rev. Lett.}
{\bf 52} (1984) 2009; A.O. Barut and M. Pav\v{s}i\v{c}: Class. Quant. Grav.
{\bf 4} (1987) L131.

\item [{[2]}\hfill] E. Whittaker: {\em A History of the Theories of
Aether and Electricity}, vol. 1 (Humanities Press; N.Y., 1973).
  
\item [{[3]}\hfill] F. Rohrlich: ``The Electron: Development of
the First Elementary Particle Theory",  in J. Mehra (ed.), {\em The
Physicists Conception of Nature}, p. 331-369 (Reidel; Dordrecht, 1973).

\item [{[4]}\hfill] M. Abraham: {\em Ann. Physik} {\bf 10} (1903) 105.

\item [{[5]}\hfill] H. A. Lorentz: {\em Theory of Electrons}, 2nd.
ed. (Dover; N.Y., 1952).

\item [{[6]}\hfill] J. D. Jackson: {\em Classical Electrodynamics}
(J. Wiley; 1962).

\item [{[7]}\hfill] P. A. M. Dirac: {\em Proc.. Roy. Soc. London}  
A{\bf 168} (1938)

\item [{[8]}\hfill] See e.g. P. Caldirola: {\em Suppl. Nuovo Cim.} {\bf 3}
(1956) 297.

\item [{[9]}\hfill] P. A. M. Dirac: {\em Proc. Roy. Soc. London}  A{\bf
209} (1951) 292; A{\bf 212} (1952) 330; A{\bf 223} (1954) 458.

\item [{[10]}\hfill] P. A. M. Dirac: {\em Nature} {\bf 168} (1951) 906. 

\item [{[11]}\hfill] Cf. A. Proca: {\em J. de Phys. et Radium} (vii) {\bf 7}
(1936) 347.  See also E. Schroedinger: {\em Proc. R. Irish Acad.} A{\bf 48}
(1943) 135.

\item [{[12]}\hfill] R. Righi and G. Venturi: {\em Int. J. Theor.
Phys.} {\bf 21} (1982) 63.

\item [{[13]}\hfill] W. A. Rodrigues Jr. and
V.L.Figueiredo: {\em Int. J. Theor. Phys.} {\bf 24} (1990) 413. 

\item [{[14]}\hfill] W. A. Rodrigues Jr., E. Recami, A. Maia Jr. and
M.A.F. Rosa: {\em Phys. Lett.} B{\bf 220} (1989) 195; \ B{\bf 173}
(1986) 233; B{\bf 188} (1987) E511. \ See also E. Mignani and E. Recami:
{\em Nuovo Cim.} A{\bf 30} (1975) 533; E. Recami: {\em Rivista Nuovo
Cim.} {\bf 9} (1986) issue no.6, pages 85, 156, and refs. therein.  

\item [{[15]}\hfill] W. A. Rodrigues Jr., M.A.  Faria-Rosa, A.  Maia
Jr. and  E. Recami:  {\em Hadronic Journal} {\bf 12} (1989) 187.

\item [{[16]}\hfill] D. Hestenes: {\em Space-Time Algebra} (Gordon \&
Breach; N.Y., 1966).

\item [{[17]}\hfill] M. Riesz: {\em Clifford Numbers and Spinors},
Lecture series no.38 (Inst. for Fluid Mech. and Appl.
Math.; Univ. of Maryland, 1958).

\item [{[18]}\hfill] R. Penrose and W. Rindler: {\em Spinors and
Space-Time}, vol. I (Cambridge Univ. Press; 1983).

\item [{[19]}\hfill] A. O. Barut: {\em Electrodynamics and Classical
Theory of Fields and Particles} (Dover; N.Y., 1980).

\item [{[20]}\hfill] N. Cabibbo and E. Ferrari: {\em Nuovo Cimento} {\bf
23} (1962) 1147.

\item [{[21]}\hfill] P. Caldirola, M. Pav\v{s}i\v{c} and E. Recami: 
{\em Nuovo Cimento}
B{\bf 48} (1978) 205; E. Recami: {\em Prog. Part. Nucl. Phys.}
{\bf 8} (1982) 401; {\em Found. Phys.} {\bf 13} (1983) 341; P. Caldirola: {\em
Nuovo Cimento} A{\bf 49} (1979) 497;
E. Recami: in {\em Old and New Questions in Physics...},
ed. by A. van der Merwe (Plenum; New York, 1982), p.377.

\item [{[22]}\hfill] D. Hestenes:  {\em Am. J. Phys.} {\bf 47} (1979) 399.

\end{list}

\end{document}